\documentclass[conference]{IEEEtran}
\IEEEoverridecommandlockouts

\usepackage{cite}
\usepackage{amsmath,amssymb,amsfonts}
\usepackage{algorithmic}
\usepackage{graphicx}
\usepackage{textcomp}
\usepackage{tabularx}
\usepackage{booktabs}
\usepackage{array}
\newcolumntype{Y}{>{\raggedright\arraybackslash}X}
\usepackage{xcolor}
\usepackage{url}
\usepackage[font=footnotesize,skip=1pt]{caption}
\usepackage{enumitem}

\setlist[itemize]{leftmargin=*,nosep,topsep=1pt,itemsep=0pt,parsep=0pt}
\setlist[enumerate]{leftmargin=*,nosep,topsep=1pt,itemsep=0pt,parsep=0pt}

\AtBeginDocument{
  \setlength{\abovedisplayskip}{3pt plus 1pt minus 1pt}
  \setlength{\belowdisplayskip}{3pt plus 1pt minus 1pt}
  \setlength{\abovedisplayshortskip}{2pt plus 1pt minus 1pt}
  \setlength{\belowdisplayshortskip}{2pt plus 1pt minus 1pt}
}

\def\BibTeX{{\rm B\kern-.05em{\sc i\kern-.025em b}\kern-.08em
    T\kern-.1667em\lower.7ex\hbox{E}\kern-.125emX}}

\begin{document}

\title{Teaching Machine Learning to Software Engineers}

\author{
\IEEEauthorblockN{1\textsuperscript{st} Nafiseh Kahani}
\IEEEauthorblockA{
\textit{Carleton University, Canada}\\
NafisehKahani@cunet.carleton.ca
}
\and
\IEEEauthorblockN{2\textsuperscript{nd} Jason Jaskolka}
\IEEEauthorblockA{
\textit{Carleton University, Canada}\\
JasonJaskolka@cunet.carleton.ca
}
}

\maketitle

\begin{abstract}
Machine learning (ML) and Artificial Intelligence (AI) components are increasingly embedded in software products, yet undergraduate software engineering (SE) curricula rarely provide systematic preparation for building, testing, deploying, and maintaining AI/ML-based software systems. This paper aims to provide evidence-based guidance for integrating AI/ML-relevant content into core SE education. We compile and define a structured inventory of topics relevant to SE practice in AI/ML-based software, then map these topics against required courses in a set of representative SE curricula to identify coverage gaps. To assess educational priorities and feasibility, we survey SE instructors on topic importance and integration constraints. Based on the crosswalk between topic definitions, curriculum coverage, and instructor prioritization, we derive a guideline that recommends where and how high-priority topics can be embedded within existing SE courses. 
\end{abstract}

\begin{IEEEkeywords}
software engineering education, machine learning, curriculum, MLOps, survey.
\end{IEEEkeywords}

\section{Introduction}

Machine learning (ML) and Artificial Intelligence (AI) are reshaping software engineering (SE) practice. Beyond programming, design, and requirements analysis, software engineers are increasingly expected to understand how AI/ML components integrate into larger systems. This includes working with data pipelines, integrating pre-trained models, and addressing cross-cutting concerns such as performance, fairness, and reliability. While AI/ML specialists often focus on designing ML algorithms and training models, software engineers need sufficient ML knowledge to collaborate effectively and support deployment in real applications.

Universities have begun to respond by integrating AI and ML into their software engineering curricula through formal minors and course offerings. For instance, Georgia Tech offers an interdisciplinary minor in Applications of AI and ML~\cite{gtAITechMinor}, Duke University provides a Minor in Machine Learning and Artificial Intelligence~\cite{duke}, and Northwestern University has established a dedicated AI minor~\cite{northwesternAIMinor}. However, these initiatives often treat AI/ML as a separate discipline rather than asking which AI/ML topics should be part of every software engineering student's education. Recent curricular guidance has emphasized this need: the \emph{Computer Science Curricula 2023 (CS2023)} report highlights AI, security, and ethics as underrepresented areas in computer science curricula~\cite{CS2023}, and the \emph{Guide to the Software Engineering Body of Knowledge (SWEBOK v4)} explicitly incorporates AI, ML, and emerging technologies into its foundational knowledge areas~\cite{SWEBOKv4}. At the same time, prior work on \emph{software engineering for machine learning} highlights recurring challenges, including data dependencies, testing and verification, technical debt, and post-deployment monitoring and debugging, which call for software engineering-tailored AI/ML competencies rather than isolated AI/ML courses~\cite{Amershi2019,Sculley2015,Breck2017,Paleyes2022}.

Despite these needs and calls, undergraduate software engineering curricula do not systematically prepare graduates to engineer and maintain software that incorporates AI/ML components. 
To address this gap, we take a two-step approach. First, we compile a broad set of ML-related topics from the literature and examples of real-world ML-based systems, any software system that uses AI and ML to perform a specific function or enhance its capabilities, organizing them into five categories: Mathematical Foundations (MF), Data Structures and Algorithms (DSA), Platform and Computing (PC), Model Development, Maintenance and Operations (MDMO), and Applications of ML in Software Engineering (AML). We then compare these topics against existing software engineering curricula to identify which are consistently covered and which are missing. Second, we survey software engineering instructors to assess the uncovered topics using a four-option scale: \emph{covered to the expected depth}, \emph{not covered but should be}, \emph{not covered and not relevant}, and \emph{not sure/unable to assess}. Based on the results, we provide recommendations for integrating the concepts and discuss the challenges in this context. This work is organized around the following research questions:
\begin{itemize}
  \item \textit{RQ1:} Which ML-related topics for software engineering are identified through synthesis of prior literature and examples of ML-based software systems, and how can they be organized into a topic framework?
  \item \textit{RQ2:} To what extent are these topics covered in required undergraduate software engineering curricula, and which topics are most consistently missing?
  \item \textit{RQ3:} How do software engineering instructors assess the uncovered topics (e.g., should be included vs not relevant vs unsure), and which topics emerge as the highest priorities for integration?
\end{itemize}

Our results show that while several MF basics (e.g., probability/distributions, gradient descent, and evaluation metrics) exhibit high coverage and support, many advanced MF topics (e.g., Maximum-Likelihood Estimation (MLE), matrix decompositions, advanced optimization) remain underrepresented. In infrastructure, core PC topics (e.g., databases, pipelines, virtual machines (VMs)/containers) are relatively well covered, whereas distributed storage, high-performance computing, and load balancing appear less consistently. Topics in MDMO and AML, including model lifecycle management (versioning, monitoring), data-centric requirements, prompt engineering, automated debugging, software security, and code quality/maintainability, receive strong support from instructors to be taught, yet have \emph{low} reported coverage across curricula. These findings motivate prioritizing applied ML-in-Software Engineering and Machine Learning Operations (MLOps) topics in undergraduate software engineering programs while incrementally closing mathematical and systems gaps.


\section{Topic Framework (RQ1)}\label{sec:concept}
This section answers \textit{RQ1} by (i) describing the sources and recency of the topic inventory, (ii) explaining how topics were consolidated and categorized, and (iii) clarifying the assumed role of a software engineer in ML-based systems, since this premise drives the scope of the framework and subsequent analyses (curriculum comparison and instructor survey).

\noindent \textit{RQ1: Which ML-related topics for software engineering are identified through synthesis of prior literature and examples of ML-based software systems, and how can they be organized into a topic framework?}

\subsection{Scope: software engineer in an ML-based system}
We define an \emph{ML-based system} as a software system whose behavior depends in part on a learned model and the associated data and deployment pipeline. In this study, \emph{software engineer} refers to practitioners who may be responsible for (a) integrating a model into a larger software system, (b) building and operating the infrastructure and pipelines that support training/deployment/monitoring, and/or (c) assuring quality attributes (testing, reliability, safety, security, compliance) of the overall system. Our premise is \emph{not} that all software engineers train models from scratch; rather, they often collaborate with ML specialists and must understand the lifecycle concerns that affect system behavior and maintenance when models and data evolve. This premise informs which topics are considered relevant and how they are grouped.

\subsection{Topic sourcing and recency}
We compiled an initial pool of candidate topics from two source types:

\textit{(S1) Literature on SE for ML.}
We conducted a systematic search over peer-reviewed research  that explicitly discusses engineering practices for ML-based systems (e.g., testing ML, deployment, MLOps, monitoring, data management). We limited research sources to 2020--2025 and focused on widely recognized venues in software engineering (ICSE-SEET, ICSE, ESEM, ESEC/FSE, and journals IEEE Transactions on Software Engineering, and IEEE Transactions on Education).  
We used keyword combinations such as \texttt{(software engineering OR testing OR maintenance) AND (machine learning OR ML-based systems OR MLOps)}. 

\textit{(S2) Practitioner descriptions and examples of ML-based systems.}
To capture operational topics that are often underrepresented in academic writing, we also extracted topics from practitioner-oriented descriptions of ML pipelines and operations, including end-to-end architectural descriptions and examples of production ML-based systems (e.g.,~\cite{equalexperts_mlops_playbook_v31_2022,datatonic_mlops_guide_2021,aws_mlops_checklist_2023,googlecloud_mlops_cd_ct_2024}). We limited this set to artifacts published between 2020-2025. These artifacts were used to cross-check and confirm that the topic inventory covers common deployment and maintenance practices (e.g., model versioning, monitoring, data drift response, and rollback strategies). We used 2020--2025 because many current SE-for-ML topics and practices are described most  in this period, including MLOps \cite{Kreuzberger2023MLOps}, monitoring, and data drift handling, and because several widely used standards and curriculum documents for AI and software engineering were published during these years (e.g., ISO/IEC 5338:2023, ISO/IEC 25059:2023, and SWEBOK V4.0). Using this window helps keep the topic inventory close to what instructors and practitioners use today.

\subsection{Topic normalization and codebook construction}
We merged extracted items into a single candidate list and then refined it through an iterative normalization process:
\begin{enumerate}
\item \textit{Deduplication and synonym consolidation:} we merged near-synonyms under canonical names (e.g., grouping ``model monitoring,'' ``production monitoring,'' and ``model observability'' under a single topic) while preserving distinctions when they imply different engineering actions (e.g., ``offline model evaluation'' vs.\ ``online monitoring and alerting'').

    \item \textit{Definition writing:} for each retained topic, we wrote a short operational definition and inclusion/exclusion notes (a lightweight codebook) to support consistent curriculum mapping.
\end{enumerate}
The final list includes 36 papers (e.g.,~\cite{warnett2024understandability,obrien202223}) for S1 and 11 resources for S2. 
We organized the topics into five categories using an inductive grouping process informed by common SE curriculum structures and ML system lifecycles. One author proposed an initial grouping based on topical affinity and the lifecycle stage(s) in which a topic is most salient. Then, another author reviewed the topic definitions and category assignments, and revised/discussed unclear boundaries or inconsistent labeling.
The resulting five categories are:
\begin{enumerate}
    \item \textit{Mathematical Foundations (MF):} mathematical concepts needed to reason about ML behavior at a basic level.
    \item \textit{Data Structures and Algorithms (DSA):} algorithmic and representational concepts often used in ML-related computation.
    \item \textit{Platform and Computing (PC):} infrastructure and systems concerns enabling scalable ML workloads and deployment.
    \item \textit{Model Development, Maintenance, and Operations (MDMO):} topics spanning evaluation, deployment, monitoring, versioning, incident response, and lifecycle maintenance of models in production.
    \item \textit{Applications of ML in Software Engineering (AML):} use of ML to support SE tasks (e.g., defect prediction, code-related analytics) and the engineering implications of adopting such tools.
\end{enumerate}

\subsection{Using curricula to focus the study’s missing topics}
Because our goal is to highlight ML-related SE topics that are \emph{not consistently addressed} in typical SE core curricula, we compared the full topic inventory against undergraduate curricula from five universities. We selected five institutions using the following criteria: (i) the institution offers an \emph{undergraduate software engineering program} (or a closely equivalent SE-focused track), (ii) the program requirements and course descriptions are publicly accessible in sufficient detail to support topic mapping, and (iii) the set collectively represents different institutional types (e.g., research-intensive vs.\ teaching-focused) rather than any single ranking-based criterion. We accessed the curricula between 2020 and 2025. To preserve neutrality, we report the institutions anonymously as U1--U5 (Table~\ref{tab:institutions}): U1 (North America, research-intensive), U2 (North America, teaching-focused), U3 (Europe, research-intensive), U4 (Asia, teaching-focused), and U5 (Australia, research-intensive).

\begin{table*}[t]
\caption{Anonymized institutions and required SE courses analyzed.}
\label{tab:institutions}
\scriptsize
\begin{tabularx}{\textwidth}{lY}
\toprule
\textbf{ID} & \textbf{Required SE courses included in the mapping} \\
\midrule
U1 & Programming \& Data Management; Foundation of Imperative Programming; Object-Oriented Software Development; Algorithms \& Data Structures; Discrete Structures I/II; Programming Languages; Computer Organization \& Architecture; Software Requirements Engineering; Real-Time Concurrent Systems; Operating Systems; Database Management Systems; Software Validation; Software Architecture \& Design; Software Economy \& Project Management; Human-Computer Interaction; Network \& Software Security. \\
U2 & Software Development \& Problem Solving I/II; Discrete Mathematics for Computing; Introduction to Software Engineering; Web Engineering; Software Process \& Project Management; Probability \& Statistics; Software Design for Computing Systems; Human-Centered Requirements \& Design; Analysis of Algorithms; Engineering Secure Software; Software System Requirements \& Architecture; Operating Systems; Database Systems. \\
U3 & Computing Fundamentals; Computing Systems; Introduction to Computational Thinking; Introduction to Object-Oriented Programming; Object-Oriented Software Engineering; Algorithmic Foundations; Networks \& Operating Systems Essentials; Algorithms \& Data Structures; Web Application Development; Algorithmics; Data Fundamentals; Human-Centred Systems Design \& Evaluation; Systems Programming; Professional Software Development. \\
U4 & Programming; Discrete Structures; Data Structures \& Algorithms; Advanced Programming; Probability \& Statistics; Computer Structure \& Language; Operating Systems; Database Systems; Computer Networks; Artificial Intelligence; System Design \& Analysis; Data \& Network Security. \\
U5 & Programming Fundamentals; Computer Systems Fundamentals; Software Engineering Fundamentals;  Discrete Mathematics; Software Construction: Techniques \& Tools; Object-Oriented Design \& Programming; Data Structures \& Algorithms; Finite Mathematics; Probability, Statistics \& Information;  Requirements \& Design of Software Systems; Software Testing \& Quality Assurance; Database Systems; Computer Networks \& Applications. \\
\bottomrule
\end{tabularx}
\end{table*}

\noindent \emph{Curriculum mapping protocol.}
We focused on required core courses and mapped course descriptions (and syllabi when available) to topic definitions in the codebook. A topic was marked \emph{covered} only when it was explicitly indicated in course content (keywords, learning outcomes, weekly topics, or detailed description) rather than inferred from broad course titles. Ambiguous cases were marked as \emph{uncertain} and resolved using the codebook definitions. Using this mapping, we filtered out topics that are consistently present in SE curricula (e.g., general programming, discrete mathematics, and baseline DSA concepts). The remaining topics, those not consistently included across the five universities, are the focus of the rest of the paper (curriculum comparison and instructor survey), because they represent the most likely gaps in current SE education for ML-based systems.

In the sections that follow, we discuss each category in detail, emphasizing the topics and their importance for future software engineers.

\subsection{Mathematical Foundations (MF)}  
Mathematical Foundations (MF) are widely recognized as essential for understanding and applying ML topics. These topics form the basis for representing data, training models, and reasoning about uncertainty. While some of these topics are covered in most undergraduate software engineering curricula, many others remain underrepresented, creating barriers to fully preparing students for ML practice~\cite{sulmont2019can}. Below, we discuss MF topics in four groups. 

\textbf{Probability and Statistics.} Topics such as \textit{basic probability rules}, \textit{common distributions} (Gaussian, Bernoulli), and \textit{statistical inference} are generally well covered in software engineering programs. However, more ML-specific areas, such as \textit{evaluation metrics} (precision, recall), the \textit{Central Limit Theorem}, and especially \textit{Bayesian networks and probabilistic models}, are often missing or taught only superficially~\cite{murphy2012machine,koller2009probabilistic,powers2011evaluation}. Advanced probabilistic reasoning, including \textit{Bayesian methods}, is also rarely covered at the undergraduate software engineering level, yet it is crucial for modern probabilistic inference~\cite{robert2004monte}.  

\textbf{Linear Algebra and Data Representation.} Core linear algebra topics such as \textit{vectors}, \textit{matrices}, and \textit{basic matrix operations} which are the core mathematical tools underlying of Deep Learning (DL) and Large Language Models (LLM) are usually taught. However, both DL and LLMs often require more advanced techniques that are not emphasized in software engineering curricula. These include \textit{orthogonal transformations and projections}, \textit{dimensionality reduction through Principal Component Analysis (PCA)}, \textit{norms and distances} (L1, L2), and \textit{matrix decompositions} (LU, QR). These topics are fundamental for tasks such as data representation, feature extraction, and optimization, but they are frequently underrepresented in existing programs~\cite{golub2013matrix,jolliffe2016principal,tibshirani1996regression,cortes1995support}.  

\textbf{Calculus and Optimization.} Standard software engineering curricula typically cover introductory calculus, including differentiation and integration. However, ML techniques, especially DL and LLMs demand more specialized knowledge. \textit{Gradient descent} is the cornerstone optimization technique for training of these models, and its connection with the \textit{chain rule and backpropagation} is essential for DL models~\cite{rumelhart1986learning,ruder2016overview}. Likewise, \textit{Maximum-Likelihood Estimation (MLE)} is a unifying framework for parameter estimation but is not often introduced in software engineering programs~\cite{casella2002statistical}. Also, advanced optimization methods—such as \textit{momentum}, \textit{Adam}, \textit{convex optimization}, and \textit{Newton or Quasi-Newton methods}—are widely used in practice, especially in training DL models, yet are seldom taught in software engineering curricula~\cite{boyd2004convex,kingma2015adam,sutskever2013importance,nocedal2006numerical}.  

\textbf{Stochastic Processes.} Finally, \textit{Markov chains} are a foundational tool for modeling sequential processes. While these appear in some computer science curricula, they are not typically included in software engineering programs. Their role in reinforcement learning, probabilistic modeling, and sequential decision-making makes them an important yet often missing component~\cite{norris1998markov}.  

In summary, while basic topics such as probability rules, simple distributions, and introductory calculus are covered in software engineering education, many  topics—such as evaluation metrics, PCA, Bayesian networks, MCMC, advanced optimization techniques, and matrix decompositions—remain underrepresented. Addressing these gaps is critical for equipping future software engineers with the mathematical background needed to understand and apply ML techniques effectively.

\subsection{Data Structures and Algorithms (DSA)}  
This category covers the data structures and algorithms essential for developing ML-based systems. As discussed in Section~\ref{sec:concept}, we focus on concepts not consistently covered in SE programs, which leads to filtering out many DSA concepts that are nevertheless essential for ML systems. We group the topics into four subcategories that reflect their particular role in ML-based systems.  

\textbf{Representations and Computation.}  
\textit{Tensors} and \textit{matrices} form the backbone of ML, providing a universal way to represent multi-dimensional data and model parameters~\cite{Goodfellow-et-al-2016}. \textit{Sparse matrices} are indispensable when working with very high-dimensional inputs, where most entries are zero, such as in text or recommendation data~\cite{Trefethen1997}. These operations are organized through \textit{computational graphs}, which describe how data flows through a model and enable automatic differentiation~\cite{Bastien2012}. This capability underpins \textit{backpropagation}~\cite{rumelhart1986learning}, allowing gradient-based training to scale to modern deep networks.  

\textbf{Efficient Data Structures.}  
Several specialized data structures enhance efficiency in ML-based systems. \textit{KD-trees} accelerate nearest-neighbor searches and clustering~\cite{Bentley1975}, \textit{tries} enable efficient handling of sequences and text~\cite{Fredkin1960}, and \textit{Bloom filters} allow high-speed, memory-efficient set membership checks at scale~\cite{Bloom1970}. Other structures such as \textit{skip lists}~\cite{Pugh1990} and \textit{priority queues}~\cite{Cormen2009} support efficient ordering and scheduling, making them valuable in streaming and online learning contexts. While these appear in general software engineering curricula, their ML-specific use emphasizes efficiency in large-scale, data-intensive computation.  

\textbf{Optimization Methods.}
\textit{Gradient descent} and its \textit{stochastic variants (SGD)} remain critical for training deep learning models~\cite{Robbins1951,Bottou2010}. Modern optimizers such as \textit{Adam}~\cite{kingma2015adam} build on these foundations. \textit{Convex optimization} offers strong guarantees for models such as logistic regression and support vector machines~\cite{boyd2004convex}. In settings where gradients are unavailable or the loss landscape is highly non-convex, heuristic strategies such as \textit{genetic algorithms}~\cite{Holland1975} provide alternatives. \textit{Matrix factorization} is a powerful tool for dimensionality reduction and collaborative filtering~\cite{Koren2009}, while \textit{autoencoders} extend this principle through neural architectures that learn compact latent spaces directly from data~\cite{Hinton2006}.  

\textbf{Advanced ML Algorithms.}  
As ML expands into big data and real-time analysis, advanced algorithms have become central. Ensemble methods such as \textit{boosting} (e.g., AdaBoost~\cite{Freund1997} and XGBoost~\cite{Chen2016}) combine multiple weak learners into highly effective models, often setting benchmarks on structured-data tasks. \textit{Sliding window} methods are crucial for real-time sequence and streaming applications~\cite{Datar2002}. Classical algorithms like \textit{flood fill}~\cite{Rosenfeld1966} remain widely used in image segmentation and spatial analysis. Although not always emphasized in software engineering, these algorithms form the computational toolkit that enables ML-based systems to operate effectively in modern, data-rich environments.

\subsection{Platform and Computing (PC)}  
This category covers the essential software and hardware infrastructure required to operate ML-based systems effectively, ensuring that models can be trained efficiently and deliver accurate predictions. We classify these topics into three key areas: Storage, Computing, and Networking.  

\textbf{Storage.}  
ML-based systems depend on efficient mechanisms for managing the massive datasets involved in training and inference. Relational databases~\cite{Codd1970} and NoSQL systems~\cite{Chang2006} provide structured and flexible access to data, while data warehouses~\cite{Inmon1996} and data lakes~\cite{Sawadogo2019} serve as repositories for analytical and raw data, respectively. Distributed storage systems such as the Google File System~\cite{Ghemawat2003} and Hadoop Distributed File System (HDFS)~\cite{Shvachko2010} ensure scalability and fault tolerance, while in-memory storage solutions like Redis~\cite{Sanfilippo2010} support low-latency access for real-time applications. To maintain reproducibility and trace data evolution, \textit{data versioning} frameworks such as Data Version Control (DVC) have become increasingly important. These storage capabilities are tied together through data pipelines, which move and transform data reliably from source to model~\cite{Gonzalez2010}.  

\textbf{Computing.}  
The computational backbone of ML relies on specialized hardware and supporting infrastructure. High-performance devices such as Graphics Processing Units (GPUs)~\cite{Nickolls2008} and Tensor Processing Units (TPUs)~\cite{Jouppi2017} enable large-scale training via massive parallelism. To manage these resources, \textit{load balancing} and \textit{cluster schedulers} (e.g., Kubernetes~\cite{Burns2016}, Borg~\cite{Verma2015}) distribute computation across nodes. Virtual machines~\cite{Goldberg1974} and containers~\cite{Merkel2014} provide portability and isolation of ML environments, while distributed frameworks such as MapReduce~\cite{Dean2008} and Spark~\cite{Zaharia2016} support large-scale training and data processing. Distributed computing framework enable the efficient execution of fundamental linear algebra operations such as LU and QR decompositions~\cite{Trefethen1997}, providing the foundation for large-scale ML workloads. Their performance is further enhanced by ML-specific compilers and runtime systems such as Accelerated Linear Algebra (XLA)~\cite{XLA2017} and Tensor Virtual Machine (TVM)~\cite{Chen2018}, which optimize execution on specialized hardware.  

\textbf{Networking.}  
Distributed ML training—particularly for large language models—requires high-speed, low-latency communication between machines. Networking for ML leverages hardware such as InfiniBand~\cite{Pfreundt2010} and optimized Ethernet solutions, as well as communication libraries like NVIDIA Collective Communications Library (NCCL)~\cite{NCCL2017} and Message Passing Interface (MPI)~\cite{MPI1994}. These frameworks are critical for synchronizing parameters, sharing data, and avoiding communication bottlenecks when training models across multiple devices and nodes.  

\subsection{Model Development, Maintenance, and Operations (MDMO)} 

As ML becomes more deeply integrated into software systems, software engineers must acquire new skills to effectively develop, deploy, and maintain these intelligent systems. Unlike traditional software development, ML-based systems introduce unique challenges at every stage of the lifecycle, from requirement gathering through design and modeling, development, deployment, and maintenance~\cite{Bosch2021EngineeringAI, Amershi2019,Sculley2015, abdelkader2024ml,kahani2020bounded,kahani2018automodel, kahani2020synthesis, bagherzadeh2020execution}.

\textbf{Requirements Analysis and Specification.}  
ML shifts the focus toward \textit{data-centric requirement gathering}, where the definition and quality of datasets become as important as specifying functionality~\cite{Ashmore2021,Serban2020, zha2025data, sambasivan2021everyone}. New \textit{quality attributes} must also be considered, including fairness, robustness, and explainability, which go beyond the traditional notions of correctness and efficiency~\cite{Arpteg2018}. Ethical and regulatory compliance further shapes the specification process, ensuring responsible development and deployment of ML-based systems~\cite{Leslie2019,Mitchell2021}.

\textbf{Design and Development.}  
Engineers must extend their expertise to include \textit{knowledge of AI/ML algorithms and libraries}, which guides the choice of suitable models, frameworks, and techniques for a given application~\cite{Zhang2020,Kim2022}. Testing and debugging are more complex than in conventional systems, as ML models rely heavily on training data and probabilistic behaviors rather than strictly deterministic logic~\cite{Breck2017}. This makes interpretability and systematic evaluation critical parts of the development process~\cite{DoshiVelez2017,Guidotti2018}.

\textbf{Deployment and Maintenance.}  
Once deployed, ML-based systems demand ongoing attention through \textit{model monitoring and debugging}, ensuring that performance remains stable in the face of changing data and environments~\cite{Breck2017,Sculley2015}. Engineers must also address the challenge of \textit{managing the lifecycle of models}, which includes retraining, updating, and versioning to maintain reliability over time~\cite{Polyzotis2018,Schelter2018, serban2024software, idowu2024machine}. Continuous maintenance in this form is essential for ML-based systems to remain accurate, trustworthy, and aligned with user and societal expectations~\cite{Amershi2019,Serban2020}.

\subsection{Applications of ML in Software Engineering (AML)} 

 Applying ML to software engineering tasks opens new ways to solve long-standing challenges, from requirements analysis to debugging and testing~\cite{Amershi2019,Allamanis2018, ajorloo2024systematic}. To make effective use of these opportunities, engineers need to understand a set of key topics that fall into three main groups: data and requirements, development, and evaluation.  

\textbf{Data and Requirements.}  
At the foundation, engineers must work with the data generated by software systems, including code repositories, execution logs, and developer activity~\cite{Hindle2012, Joshi2024PRPrediction}. This requires skills in \textit{data collection and processing}, as well as the ability to adapt ML techniques to this kind of domain-specific data~\cite{Zhang2020}. With the emergence of large language models, \textit{prompt engineering} and \textit{LLM-based code generation} have become central to automating and supporting software tasks~\cite{Chen2021Codex,FinnieAnsley2022, barke2023grounded}. ML also supports \textit{requirements engineering}, where intelligent models can help refine specifications and capture quality attributes such as fairness, robustness, and explainability~\cite{Arpteg2018,Ashmore2021}.  

\textbf{Development.}  
In the development phase, ML methods directly improve software quality and reliability. AI and ML techniques support \textit{code quality assessment and maintainability}~\cite{Allamanis2018, cabral2024investigating, Sato2019}, enforce consistency through \textit{code linting}~\cite{Svyatkovskiy2020, Holden2024CodeLinting}, and enhance \textit{software security} by identifying vulnerabilities and supporting defense mechanisms against attacks~\cite{Ejikeme2024DDoS, Ghaffarian2017, Kahani2018ReactiveDefense, Fallah2014TDPF}. ML-based tools further contribute to \textit{bug detection and automated debugging}, reducing the effort required for troubleshooting~\cite{Just2014,Pradel2018}. Beyond these, \textit{software testing optimization} uses ML to design efficient, adaptive test suites that reduce costs while maintaining strong coverage~\cite{Pan2019}.  

\textbf{Evaluation.}  
Finally, systematic \textit{evaluation and benchmarking} are necessary to measure the effectiveness of ML in software engineering. Since traditional ML evaluation metrics may not fully capture software quality, domain-specific benchmarks must be developed to assess improvements in maintainability, reliability, and overall performance~\cite{Breck2017}. Proper evaluation ensures that ML integration is both meaningful and beneficial across the entire software lifecycle.

\section{Study Design and Findings (RQ2--RQ3)}\label{sec:Survey}
Here, we describe the survey design and data collection process and present the resulting findings. This section answers \textit{RQ2} and \textit{RQ3} using instructor-reported information from their institutions: \textit{RQ2} concerns the current \emph{coverage} of topics in required SE curricula, and \textit{RQ3} concerns instructors' judgments about whether uncovered topics \emph{should be included} and which topics appear to be the highest priorities for integration.

\subsection{Design}
Building on the topic framework in Section~\ref{sec:concept}, we designed a survey to answer \textit{RQ2} and \textit{RQ3}. For \textit{RQ2}, we asked instructors to report whether each topic is covered in their required SE curriculum \emph{to the expected depth}. For topics reported as not covered, \textit{RQ3} captures instructors' judgments about whether the topic should be included, is not relevant, or is difficult to assess. This design is particularly important for emerging areas such as ML in software engineering, where expert input helps balance disciplinary needs with pedagogical feasibility.

The survey was structured around the five categories introduced earlier: Mathematical Foundations (MF), Data Structures and Algorithms (DSA), Platform and Computing (PC), Model Development, Maintenance and Operations (MDMO), and Applications of ML in Software Engineering (AML). To support analysis and reporting, each topic within these categories was assigned a unique identifier (e.g., A1, B4, C7). The complete list of topics and identifiers is shown in Table~\ref{tab:all-concepts}; we use these identifiers throughout the results for clarity and conciseness.

For each topic, respondents selected one of four categorical options: \emph{covered to the expected depth}, \emph{not covered but should be}, \emph{not covered and not relevant}, and \emph{not sure/unable to assess}. This format captures both current coverage (RQ2) and perceived educational importance (RQ3). In addition, open-text questions invited instructors to explain their reasoning, identify barriers to integration, and suggest supports that could make adoption feasible. A final consent option allowed participants to decide whether their responses could be included in the aggregate analysis.

To improve clarity and reduce ambiguity, we piloted the survey with four colleagues. Two had recently taught machine learning courses and provided feedback from both teaching and subject-matter perspectives. A third colleague had taught first- and second-year software engineering courses, and a fourth had taught third- and fourth-year software engineering courses.  Based on their input, we refined and clarified the wording of the survey items to improve readability and promote consistent interpretation.

\begin{table*}[t!]
\caption{Survey Topics Across Categories A--E}
\label{tab:all-concepts}
\small
\begin{tabularx}{\textwidth}{lY lY}
\toprule
\textbf{ID} & \textbf{Topic} & \textbf{ID} & \textbf{Topic} \\
\midrule
\multicolumn{4}{l}{\textbf{Category A: Mathematical Foundations}} \\
A1  & Advanced Bayesian methods (MCMC)          & A2  & Advanced optimization (Newton) \\
A3  & Basic probability rules                   & A4  & Bayesian networks \& probabilistic models \\
A5  & Central Limit Theorem                     & A6  & Chain rule / back-propagation \\
A7  & Common distributions (e.g., Gaussian)     & A8  & Convex optimization basics \\
A9  & Dimensionality reduction (PCA)            & A10 & Evaluation metrics (e.g., precision, recall) \\
A11 & Gradient descent                          & A12 & Markov chains \\
A13 & Matrix decompositions (LU, QR)            & A14 & Maximum-Likelihood Estimation (MLE) \\
A15 & Norms \& distances (L1, L2, …)            & A16 & Optimization techniques (momentum, Adam) \\
A17 & Orthogonal transformations / projections  & A18 & Statistical inference \\
\midrule

\multicolumn{4}{l}{\textbf{Category B: Data Structures \& Algorithms}} \\
B1  & Tensors \& matrices                       & B2  & Computational graph \\
B3  & Data buffers / dataframes                 & B4  & Gradient-descent and variants (SGD, Adam) \\
B5  & Back-propagation                          & B6  & KD-Trees \\
B7  & Trie structures                           & B8  & Bloom filters \\
B9  & Convex optimization algorithms            & B10 & Matrix factorization \\
B11 & Skip lists \& priority queues             & B12 & Autoencoders \\
B13 & Boosting algorithms (XGBoost, AdaBoost)   & B14 & Genetic algorithms, Sliding Window, and Floodfill  \\

\midrule

\multicolumn{4}{l}{\textbf{Category C: Platform \& Computing}} \\
C1  & Relational \& NoSQL databases             & C2  & Distributed storage systems \\
C3  & Data pipelines                            & C4  & High-performance computing (GPU, TPU) \\
C5  & Load balancing                            & C6  & Data warehouses \\
C7  & In-memory data storage                    & C8  & Data versioning \\
C9  & Resource scheduling                       & C10 & Virtual machines \& containers \\
C11 & Data lakes                                & C12 & Distributed-computing frameworks (e.g., MapReduce) \\
C13 & ML-specific compilers                     & C14 & Networking for ML \\

\midrule

\multicolumn{4}{l}{\textbf{Category D: Development, Maintenance \& Operations}} \\
D1  & Data-centric requirement gathering        & D2  & New quality attributes specific to ML \\
D3  & Ethical \& regulatory compliance          & D4  & AI/ML algorithm \& library knowledge \\
D5  & Model monitoring \& debugging             & D6  & Managing the lifecycle of models \\

\midrule

\multicolumn{4}{l}{\textbf{Category E: Applications of ML in Software Engineering}} \\
E1  & Data collection / processing              & E2  & Prompt engineering \\
E3  & Evaluation \& benchmarking                & E4  & Bug detection \& automated debugging \\
E5  & Software-testing optimization             & E6  & AI/ML in requirements engineering \\
E7  & Code quality \& maintainability           & E8  & Code linting using AI/ML \\
E9  & Software security using AI/ML             &     &   \\

\bottomrule
\end{tabularx}
\end{table*}

From the survey responses, we extract three outcomes: (1) instructor-reported coverage patterns across the topic framework (RQ2), (2) instructor judgments and priorities for uncovered topics (RQ3), and (3) qualitative insights into barriers and supports for integration.

The study was reviewed and approved by the Carleton University Research Ethics Board. Participation was voluntary, and responses were collected anonymously; no personally identifying information was retained. The survey included a mix of multiple-choice items, Likert-scale questions, and short open-ended questions. 

\subsection{Collection}
After receiving ethical approval, we distributed an online survey link through social media channels, including LinkedIn and X. Participants were required to (i) have experience teaching software engineering courses and (ii) have prior knowledge of basic AI and ML.

The survey was open for 60 days, from July 11, 2025 to September 9, 2025, and yielded 48 responses. After validation, we excluded seven incomplete responses and five responses in which participants selected \emph{not sure/unable to assess} for more than 50\% of the topics. This resulted in 36 valid responses.

Responses came from 36 universities: 13 from North America, 11 from Asia, 9 from Europe, and 3 from Australia. Although the sample size is modest, it is geographically diverse and provides a useful basis for exploratory analysis.

\subsection{Results}
This section reports the survey findings for each topic category. Because the survey includes many items, we use a two-stage interpretation procedure to move beyond raw percentages and highlight patterns relevant to curriculum design.

\subsubsection{RQ2: Reported coverage of topics in required SE curricula}
For \textit{RQ2}, we operationalize \emph{coverage} as the proportion of respondents selecting \emph{covered to the expected depth} for a topic. We report coverage using three bands:
\begin{itemize}
    \item Low coverage: $<$ 25\%; Moderate coverage: 25--50\%; and High coverage: $>$ 50\%  
\end{itemize}

\subsubsection{RQ3: Instructor support and priorities for uncovered topics}
For \textit{RQ3}, we interpret both \emph{covered to the expected depth} and \emph{not covered but should be} as indicators of \emph{support for inclusion}. We interpret \emph{not covered and not relevant} and \emph{not sure/unable to assess} as indicators of \emph{low support}. Combining coverage and support yields four groups:

\begin{itemize}
    \item \emph{Low coverage/low support:} coverage $<$ 25\% and more than 50\% of responses indicate low support. These topics are rarely taught and are not broadly prioritized for inclusion.
    \item \emph{Low coverage/high support:} coverage $<$ 25\% and at least 50\% of responses indicate support for inclusion. These topics are widely viewed as important but are not currently taught.
    \item \emph{Moderate coverage/high support:} coverage 25--50\% and at least 50\% of responses indicate support for inclusion. These topics are viewed as important but are not yet widely taught.
    \item \emph{High coverage/high support:} coverage $>$ 50\% and at least 50\% of responses indicate support for inclusion. The topics are widely recognized foundations that are broadly taught.
\end{itemize}

The combined grouping provides a structured basis for identifying consensus areas, curricular gaps, and uneven adoption across institutions.

\subsubsection{Mathematical Foundations (MF)}
\begin{figure}
    \centering
    \includegraphics[width=1\linewidth]{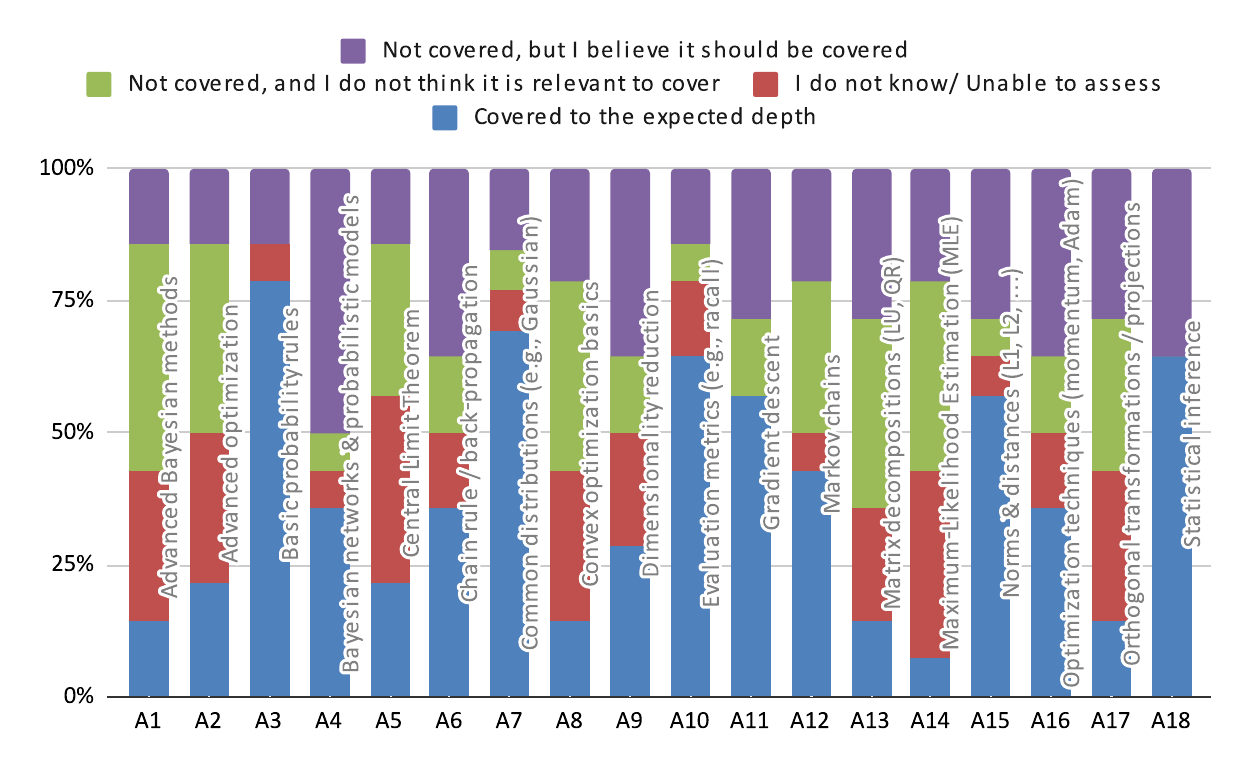}
    \caption{Distribution of responses for Mathematical Foundations (MF).}
    \label{fig:mf_result}
\end{figure}

Figure~\ref{fig:mf_result} summarizes responses for MF topics.

\textbf{RQ2 (Coverage).}
Core MF topics show high reported coverage across institutions, while several advanced topics appear less consistently covered.

\textbf{RQ3 (Support/Priority).}
\textbf{Low Coverage/Low Support.}
The following topics fall into the low coverage and low support group: Convex optimization basics, Matrix decompositions (LU, QR), Advanced Bayesian methods, Orthogonal transformations/projection, Central Limit Theorem, Advanced optimization (e.g., Newton's method), and Maximum-Likelihood Estimation (MLE). These topics are rarely reported as covered, and many respondents rated them as not relevant or were unsure.

\textbf{Low Coverage/High Support.}
No topics are categorized in this group.

\textbf{Moderate Coverage/High Support.}
Several applied or advanced methods, such as Bayesian networks \& probabilistic models, Chain rule and back-propagation, Dimensionality reduction, Markov chains, and Optimization techniques (e.g., momentum, Adam), show moderate coverage, with many remaining respondents indicating they ``should be covered.'' These are candidates for targeted integration, although responses show variation rather than full consensus.

\textbf{High Coverage/High Support.}
Core MF topics including Norms \& distances (e.g., L1, L2), Gradient descent, Evaluation metrics (e.g., precision, recall), Common distributions (e.g., Gaussian), and Basic probability rules show high coverage and strong support. Their absence in some curricula may disadvantage students relative to peers in programs where these foundations are systematically taught.

\textbf{Summary.}
Overall, respondents indicate that foundational MF topics are commonly covered, while more advanced MF topics receive weaker support in an SE-oriented context. Moderately supported topics (e.g., Bayesian models and Markov chains) appear as gaps and may warrant additional attention depending on curricular goals and constraints.

\subsubsection{Data Structures and Algorithms (DSA)}
\begin{figure}
    \centering
    \includegraphics[width=1\linewidth]{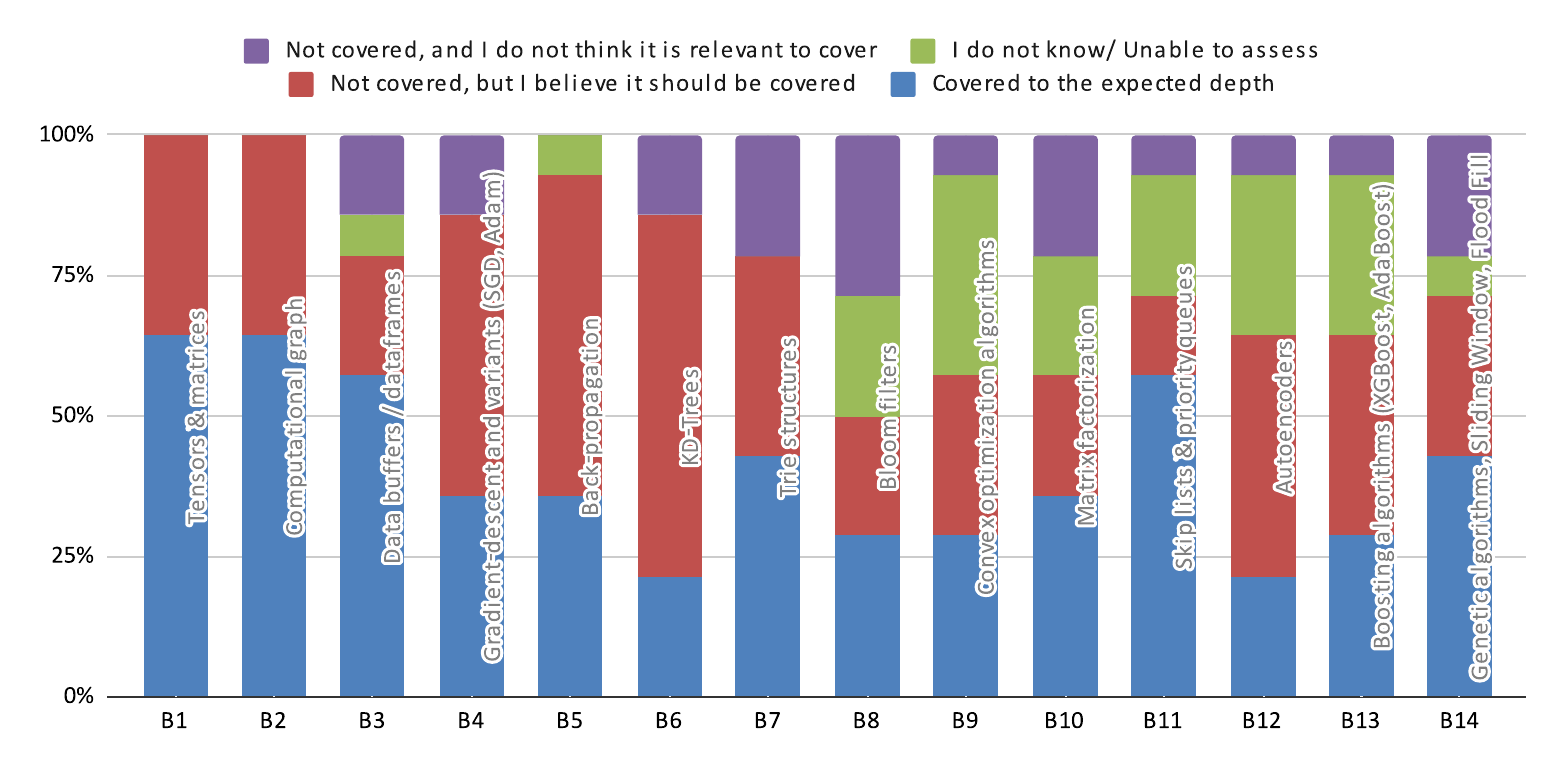}
    \caption{Distribution of responses for Data Structures and Algorithms (DSA).}
    \label{fig:dsa}
\end{figure}

Figure~\ref{fig:dsa} summarizes responses for DSA topics.

\textbf{RQ2 (Coverage).}
Most DSA topics show moderate to high reported coverage, with a small number of lower-coverage topics.

\textbf{RQ3 (Support/Priority).}
Since no items fall into the low coverage/low support group, we focus on the remaining groups.

\textbf{Low Coverage/High Support.}
Among the surveyed DSA topics, KD-tree and Autoencoders show low coverage while still being considered relevant for inclusion by respondents.

\textbf{Moderate Coverage/High Support.}
Many DSA topics, including Back-propagation, Gradient descent and its variants (SGD, Adam), Trie structures, Bloom filters, Convex optimization algorithms, Matrix factorization, Boosting algorithms (XGBoost, AdaBoost), Genetic algorithms, Sliding Window, and Flood Fill, are viewed as important to include, but their coverage remains modest (average coverage approximately 33\%). This indicates uneven adoption across programs.

\textbf{High Coverage/High Support.}
The remaining topics, including Tensors \& matrices, Computational graph, Data buffers/dataframes, and Skip lists \& priority queues, receive both high support and high coverage. However, coverage rates (57--64\%) still suggest inconsistency across universities.

\textbf{Summary.}
Respondents show broad support for DSA-related topics, but coverage is uneven across institutions, leaving students inconsistently prepared. Low-coverage topics such as KD-tree may require targeted attention to reduce gaps.

\subsubsection{Platform and Computing (PC)}
\begin{figure}
    \centering
    \includegraphics[width=1\linewidth]{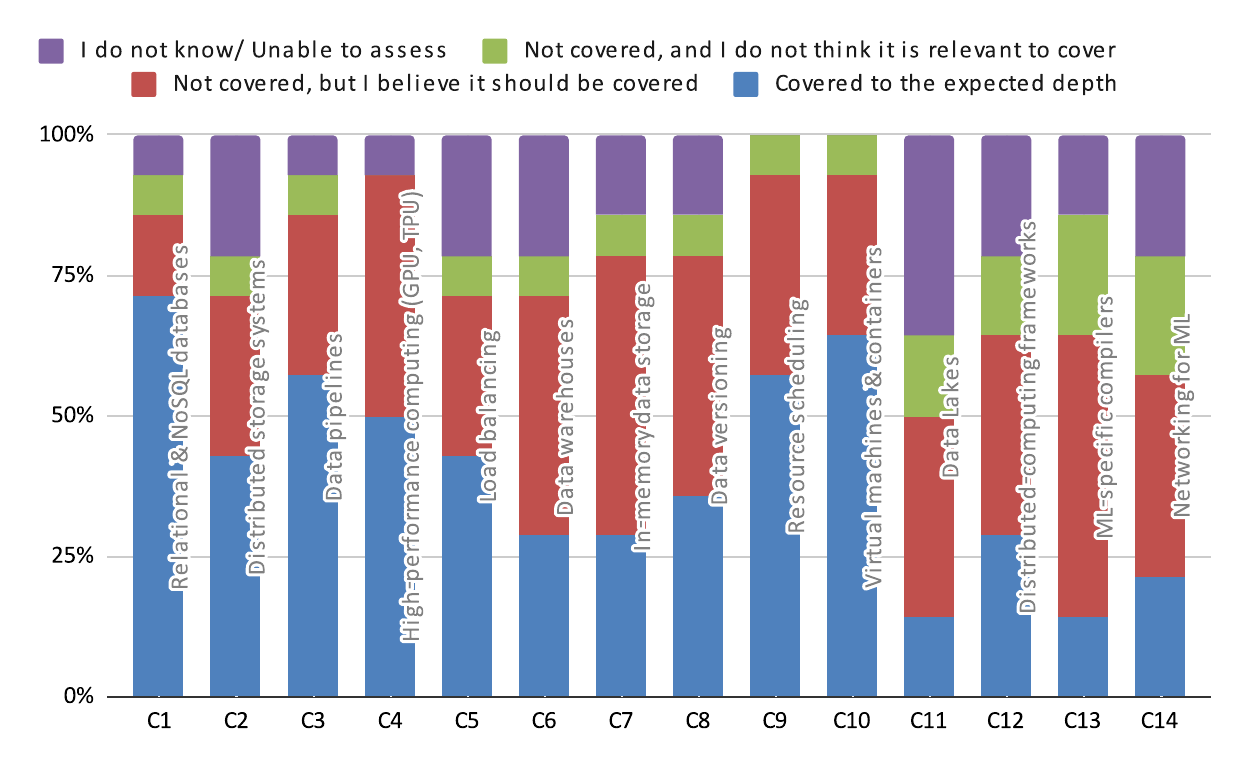}
    \caption{Distribution of responses for Platform and Computing (PC).}
    \label{fig:c}
\end{figure}

Figure~\ref{fig:c} summarizes responses for PC topics.

\textbf{RQ2 (Coverage).}
Several core infrastructure topics show high coverage, while others appear only moderately covered across institutions.

\textbf{RQ3 (Support/Priority).}
Since no items fall into the low coverage/high support group, we discuss the remaining groups.

\textbf{Low Coverage/Low Support.}
Data lakes, ML-specific compilers, and networking for ML exhibit low coverage and limited support for inclusion. These topics may be lower priority in the near term, and some (especially networking) may require substantial prerequisites.

\textbf{Moderate Coverage/High Support.}
Distributed storage systems, high-performance computing, load balancing, and data warehouses show moderate coverage but high support. The proportion of \emph{not sure/unable to assess} responses is relatively high for some of these items, suggesting a need to clarify learning objectives and provide concrete curricular examples.

\textbf{High Coverage/High Support.}
Relational/NoSQL databases, data pipelines, resource scheduling, virtual machines, and containers are well covered and broadly viewed as necessary. Nevertheless, some programs still report limited exposure, indicating room to improve consistency.

\textbf{Summary.}
Core PC topics (databases, pipelines, containers/VMs) are well supported and should remain central. Distributed storage, HPC, load balancing, and data warehouses show gaps between support and coverage, warranting greater emphasis. By contrast, data lakes, ML-specific compilers, and networking for ML appear lower priority given limited support and prerequisite demands.

\subsubsection{Model Development, Maintenance and Operations (MDMO)}
\begin{figure}
    \centering
    \includegraphics[width=1\linewidth]{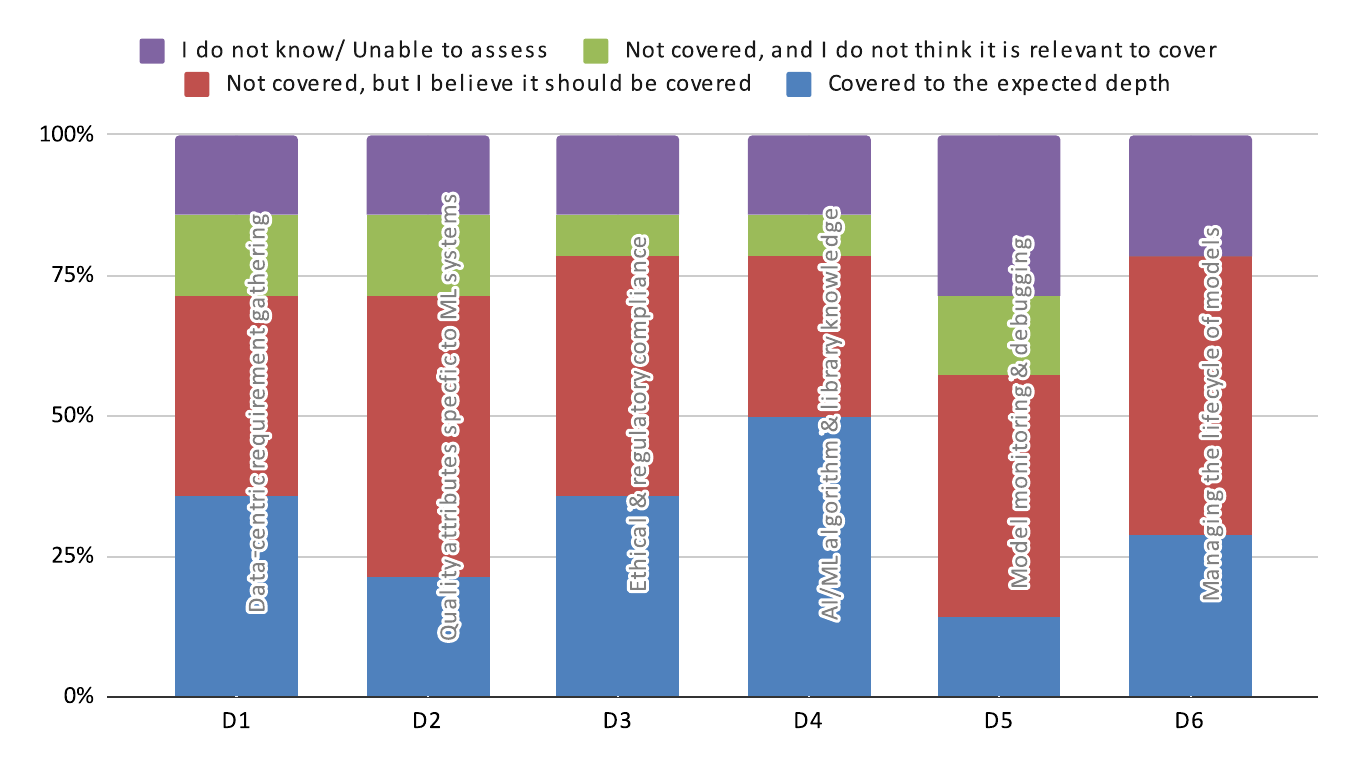}
    \caption{Distribution of responses for Model Development, Maintenance and Operations (MDMO).}
    \label{fig:d}
\end{figure}

Figure~\ref{fig:d} summarizes responses for MDMO topics.

\textbf{RQ2 (Coverage).}
Across MDMO, respondents report low-to-moderate coverage, indicating that lifecycle and operations topics are not yet consistently embedded in required curricula.

\textbf{RQ3 (Support/Priority).}
All topics in this category receive strong support, and none falls into the high coverage groups; we therefore focus on the two high-support groups.

\textbf{Low Coverage/High Support.}
Quality attributes specific to ML-based systems and model monitoring \& debugging show notably low coverage, yet both are strongly supported for inclusion. These topics are critical for successfully deploying ML-based systems and represent high-priority gaps.

\textbf{Moderate Coverage/High Support.}
Other topics in this category receive moderate coverage but are also strongly supported, particularly model lifecycle management and the introduction of new quality metrics for ML-based systems.

\textbf{Summary.}
MDMO topics are broadly viewed as important but remain insufficiently covered. The mismatch between strong support and limited coverage indicates a critical curricular gap that warrants further attention.

\subsubsection{Applications of ML in Software Engineering (AML)}
\begin{figure}
    \centering
    \includegraphics[width=1\linewidth]{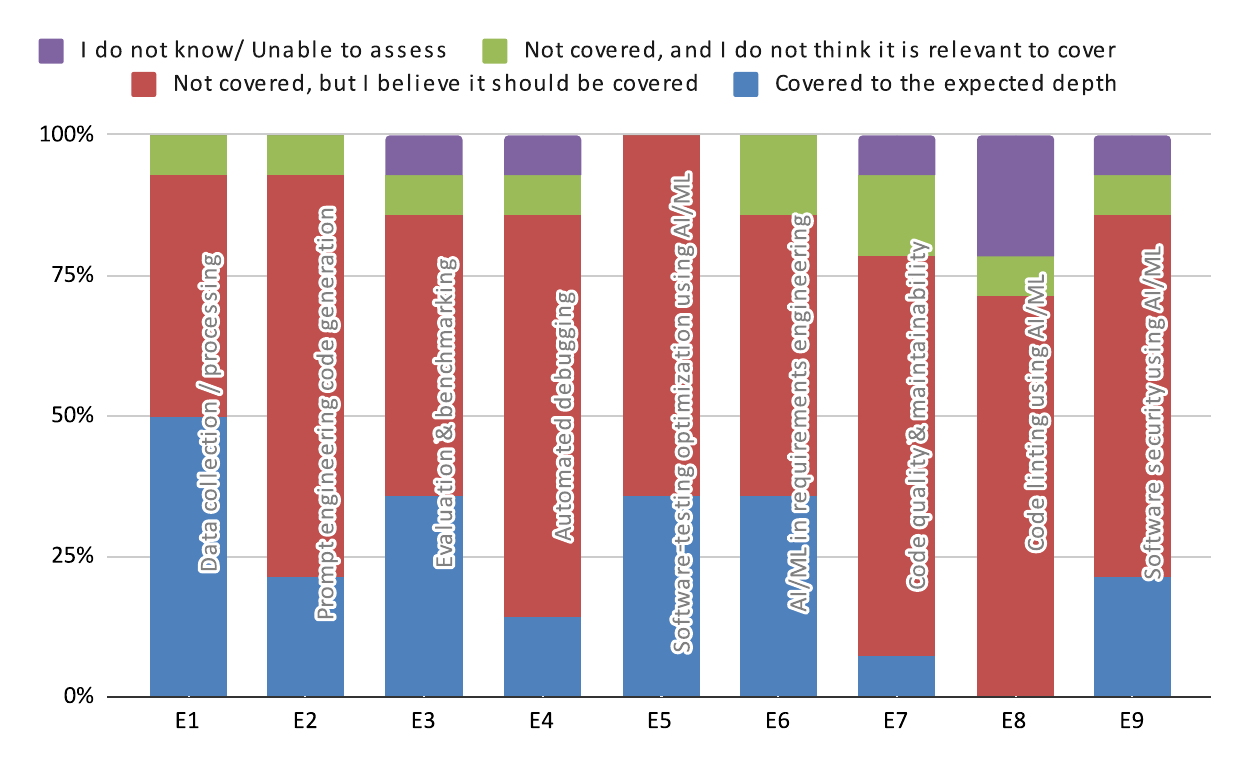}
    \caption{Distribution of responses for Applications of ML in Software Engineering (AML).}
    \label{fig:e}
\end{figure}

Figure~\ref{fig:e} summarizes responses for AML topics. All topics in this category receive strong support.

\textbf{RQ2 (Coverage).}
Reported coverage for AML topics is consistently low-to-moderate, suggesting that applied ML-in-SE competencies are not yet widely integrated into required curricula.

\textbf{RQ3 (Support/Priority).}
Because none of the topics falls into the low support groups, we focus on the two high-support groups.

\textbf{Low Coverage/High Support.}
Prompt engineering, code quality and maintainability, software security, code linting, and automated debugging show low coverage despite strong support for inclusion. Their limited presence points to high-priority gaps.

\textbf{Moderate Coverage/High Support.}
The remaining topics achieve only moderate coverage but are also strongly supported, indicating a need to expand curricular focus on applied ML within software engineering.

\textbf{Summary.}
AML topics show strong consensus in perceived value but consistently low coverage. Addressing this gap may require either dedicated offerings or systematic integration into existing SE courses.

\subsection{RQ3 Qualitative Findings}
To complement the quantitative RQ3 results, we analyzed responses to an open-text question asking participants to describe the challenges and barriers they face in teaching ML-related topics to software engineering students. We grouped responses into three recurring themes: expertise and knowledge gaps, overloaded programs, and timing, computational resources, and curriculum-change processes.

\subsubsection{Expertise and Knowledge Gaps}
Participants emphasized that teaching many of the identified topics requires expertise spanning both ML and software engineering. Several noted that their institutions lack instructors with sufficient background in both areas, which hinders effective integration and delivery. Participants also pointed to limited availability of teaching materials, best practices, and pedagogical guidance for these topics, making adoption difficult to start and sustain.

\textbf{Implication for curriculum design:} Institutions may mitigate expertise constraints via team-teaching arrangements (e.g., ML specialists co-delivering with software engineering faculty) and faculty development opportunities such as targeted workshops, summer schools, or industry partnerships.

\subsubsection{Overloaded Programs}
Several participants highlighted that SE curricula are already overloaded, and instructors struggle to cover existing core content. Finding room for additional ML-related topics therefore requires trade-offs. Some participants noted that adding ML content could displace essential SE topics, potentially weakening foundational SE preparation.

\textbf{Implication for curriculum design:} Modular integration may reduce disruption by embedding ML-related examples, labs, or assignments into existing SE courses (e.g., illustrating testing with ML-based bug prediction tools). Elective modules, minors, and capstone projects can provide deeper exposure without displacing core SE content. Curriculum mapping and prioritization based on instructor consensus can further ensure that additions are targeted and feasible.

\subsubsection{Timing, Computational Resources, and Curriculum-Change Processes}
Participants noted that revising courses and curricula to integrate new topics is often bureaucratic and time-consuming, with meaningful updates taking years. They also emphasized the time and financial costs of preparing new teaching materials, developing expertise, and adapting course structures. In parallel, several participants mentioned the need for adequate computational infrastructure (e.g., GPUs) to support hands-on assignments.

\textbf{Implication for curriculum design:} Incremental adoption strategies (e.g., optional labs or assignments) can enable near-term introduction of ML-related topics while broader curriculum revisions are under review. Sustained change may require institutional support mechanisms, including funding or recognition for developing teaching materials and shared resources (e.g., cross-institution content sharing) to distribute workload and accelerate adoption.

\section{Related Work}\label{sec:RelatedWork}
The related work is organized around the following bodies of literature:

\paragraph{Curricular guidance and competency frameworks}
While CS2023 and SWEBOK v4 identify intersections such as AI for SE and SE for AI/ML systems, they define \emph{what} knowledge areas matter without providing (i) a reusable topic inventory mappable to existing SE courses, or (ii) empirical evidence about current coverage and educator priorities. Our study complements these efforts by operationalizing SE-relevant ML/LLM competencies as a topic inventory connected to reported coverage and instructor prioritization.

\paragraph{Software engineering for ML-based systems and lifecycle concerns}
A substantial body of work highlights distinct engineering challenges of ML-based systems, including hidden technical debt, data dependencies, monitoring needs, and post-deployment maintenance~\cite{Sculley2015,Amershi2019,Breck2017,Paleyes2022}. This literature motivates incorporating lifecycle-oriented competencies (monitoring, debugging, versioning, rollback) and ML-specific quality attributes into SE education, but typically articulates challenges or practices without mapping them to current SE curricula or assessing educator consensus. Our work bridges this gap by translating lifecycle challenges into teachable topics and empirically identifying where they are undercovered yet strongly supported.

\paragraph{Teaching AI/ML at the intersection with software engineering}
Several studies describe course designs that connect ML and SE: K{\"a}stner and Kang report a course on software engineering for AI-based systems~\cite{kastner2020teaching}, and Chenoweth and Linos propose teaching ML as part of agile SE~\cite{ChenowethLinos2024}. These contributions provide detailed pedagogical designs but are typically anchored to a single course/institution and do not provide a cross-program picture of coverage gaps. In contrast, our study is curriculum-oriented, using instructor input to identify missing and prioritized topics across programs.

\paragraph{LLMs and AI tool use in software engineering education}
Recent work discusses opportunities and risks of integrating LLMs in SE education~\cite{KhanAkbarKasurinen2025LLMSEd} and provides prompt catalogs mapped to SE knowledge areas~\cite{PereiraLGA2024OpenSourceLLMsCSEET}. Emerging SE research also treats prompts as software artifacts and studies systematic prompt development practices (\emph{promptware engineering})~\cite{ChenEtAl2025PromptwareEngineering,VillamizarEtAl2025PromptsArtifacts}. Accordingly, we treat prompt-related topics as SE-facing competencies for working with LLM-based tools and components, and measure both their reported coverage and educator support.

Prior work provides high-level guidance on what AI/ML topics matter, engineering evidence that ML-based systems require lifecycle and quality practices, and course-level pedagogical designs. Our contribution connects these threads by offering an SE-focused topic inventory mappable to curricula and empirically identifying coverage gaps and instructor priorities for evidence-based integration decisions.

\section{Threats to Validity}\label{sec:Threats}
\textbf{Sample size and generalization.} The 5 institutions and 36 instructor responses form a purposive, exploratory sample selected to span institutional types and regions; the results should be read as tendencies and hypotheses for larger-scale replication rather than as statistically generalizable claims. \textbf{Self-selection and response bias.} Voluntary participation may favor instructors with an existing interest in ML or curricular reform, and self-reported views may diverge from actual classroom practice or institutional priorities. \textbf{Curriculum interpretation.} Curriculum mapping relies on public catalogs, syllabi, and program websites, which may omit instructor-level content choices and thus under- or over-estimate coverage of specific topics. \textbf{Coding subjectivity.} Topic classification is interpretive; to mitigate this we used a codebook with operational definitions and inclusion/exclusion notes, applied consistency checks across coders, and release the codebook as supplementary material to support independent replication and re-analysis.

\section{Conclusion}\label{sec:Conclusion}
This study mapped ML topics relevant to SE students across five categories and assessed their coverage through curriculum analysis and an instructor survey. Foundational topics are well covered, but advanced methods, platform-oriented competencies, and applied ML-in-SE areas remain underrepresented, with instructors citing barriers such as limited expertise, overloaded curricula, and slow revision processes. We recommend prioritizing applied and lifecycle-management topics through incremental, modular integration. Future work should develop reusable course modules and validate these findings with larger, more diverse samples.\\

\noindent\textbf{Acknowledgment:} The work was supported by the Natural Sciences and Engineering Research Council of Canada.


\def\IEEEbibitemsep{0pt plus .2pt}
\renewcommand{\baselinestretch}{0.97}\selectfont
{\footnotesize
\bibliographystyle{IEEEtran}
\bibliography{references}}

@inproceedings{kastner2020teaching,
  title={Teaching Software Engineering for AI-enabled Systems},
  author={K{\"a}stner, Christian and others},
  booktitle={Soft. Eng. Education and Training},
  pages={45--48},
  year={2020}
}

@techreport{CS2023,
  title        = {Computer Science Curricula 2023 (CS2023): The Final Report},
  author       = {{Joint Task Force on Computer Science Curricula}},
  institution  = {ACM, IEEE Computer Society, and AAAI},
  year         = {2024},
  doi          = {10.1145/3626253.3633405},
  url          = {https://ieeecs-media.computer.org/media/education/reports/CS2023.pdf}
}

@online{SWEBOKv4,
  title   = {SWEBOK V4.0: Guide to the Software Engineering Body of Knowledge},
  author  = {{IEEE Computer Society}},
  year    = {2024},
  url     = {https://www.computer.org/education/bodies-of-knowledge/software-engineering/v4/} 
}

@article{Holden2024CodeLinting,
  author  = {Holden, Darren and others},
  title   = {Code Linting using Language Models},
  journal = {arXiv preprint arXiv:2406.19508},
  year    = {2024}
}

@article{Fallah2014TDPF,
  author  = {Fallah, Mehran S. and others},
  title   = {{TDPF}: A Traceback-Based Distributed Packet Filter to Mitigate Spoofed {DDoS} Attacks},
  journal = {Security and Communication Networks},
  volume  = {7},
  number  = {2},
  pages   = {245--264},
  year    = {2014},
  doi     = {10.1002/sec.725}
}

@article{bagherzadeh2020execution,
  title={Execution of partial state machine models},
  author={Bagherzadeh, Mojtaba and others},
  journal={TSE},
  volume={48},
  number={3},
  pages={951--972},
  year={2020}
}

@inproceedings{Kahani2018ReactiveDefense,
  author    = {Kahani, Nafiseh and others},
  title     = {A Reactive Defense Against Bandwidth Attacks Using Learning Automata},
  booktitle = {Availability, Reliability and Secu.},
  year      = {2018},
  pages     = {1--6}
}

@inproceedings{Joshi2024PRPrediction,
  author    = {Joshi, Rinkesh and others},
  title     = {Comparative Study of Reinforcement Learning in {GitHub} Pull Request Outcome Predictions},
  booktitle = {SANER},
  year      = {2024},
  pages     = {489--500}}

@inproceedings{Amershi2019,
  author    = {Amershi, Saleema and others},
  title     = {Software Engineering for Machine Learning: A Case Study},
  booktitle = {ICSE-SEIP},
  year      = {2020},
  pages     = {291--300},
  doi       = {10.1109/ICSE-SEIP.2019.00042}
}

@inproceedings{Ejikeme2024DDoS,
  author    = {Ejikeme, Chibuike and others},
  title     = {Optimizing {DDoS} Detection with Time Series Transformers},
  booktitle = {Collaborative Advances in Software and COmputiNg},
  year      = {2024},
  pages     = {1--6},
  publisher = {IEEE},
  doi       = {10.1109/CASCON62161.2024.10837833}
}

@article{Sato2019,
title={On testing machine learning programs},
  author={Braiek, Houssem Ben and Khomh, Foutse},
  journal={J. of Systems and Software},
  volume={164},
  pages={110542},
  year={2020},
  publisher={Elsevier}
}

@article{Paleyes2022,
  author  = {Andrei Paleyes and others},
  title   = {Challenges in Deploying Machine Learning: A Survey of Case Studies},
  journal = {ACM Computing Surveys},
  year    = {2022},
  volume  = {55},
  number  = {6}
}

@article{ajorloo2024systematic,
  title={A Systematic Review of Machine Learning Methods in Software Testing},
  author={Ajorloo, Sedighe and others},
  journal={Applied Soft Computing},
  volume={162},
  pages={111805},
  year={2024}
}

@article{barke2023grounded,
  title={Grounded Copilot: How Programmers Interact with Code-generating Models},
  author={Barke, Shraddha and others},
  journal={Programming Languages},
  volume={7},
  number={OOPSLA1},
  pages={85--111},
  year={2023}}

@inproceedings{sambasivan2021everyone,
  title={“Everyone Wants to do the Model Work, not the Data Work”: Data Cascades in High-Stakes AI},
  author={Sambasivan, Nithya and others},
  booktitle={2021 CHI  Conf. on Human Factors in Computing Systems},
  pages={1--15},
  year={2021}
}

@book{murphy2012machine,
  title={Machine Learning: A Probabilistic Perspective},
  author={Murphy, Kevin P.},
  year={2012},
  publisher={MIT Press}
}

@book{koller2009probabilistic,
  title={Probabilistic Graphical Models: Principles and Techniques},
  author={Koller, Daphne and Friedman, Nir},
  year={2009},
  publisher={MIT Press}
}

@article{powers2011evaluation,
 title={Evaluation: from precision, recall and F-measure to {ROC}, informedness, markedness and correlation},
  author={Powers, David MW},
  journal={\url{arXiv:2010.16061}},
  year={2020}
}

@book{robert2004monte,
  title={Monte Carlo Statistical Methods},
  author={Robert, Christian P. and Casella, George},
  year={1999},
  edition={2},
  publisher={Springer}
}

@book{golub2013matrix,
  title={Matrix Computations},
  author={Golub, Gene H. and Van Loan, Charles F.},
  year={2013},
  edition={4},
  publisher={Johns Hopkins University Press}
}

@book{jolliffe2016principal,
  title={Principal Component Analysis},
  author={Jolliffe, Ian T. and Cadima, Jorge},
  year={2016},
  publisher={Springer}
}

@article{tibshirani1996regression,
  title={Regression Shrinkage and Selection via the Lasso},
  author={Tibshirani, Robert},
  journal={Methodological},
  volume={58},
  number={1},
  pages={267--288},
  year={1996}
}

@inproceedings{cortes1995support,
  title={Support-Vector Networks},
  author={Cortes, Corinna and Vapnik, Vladimir},
  booktitle={Machine Learning},
  volume={20},
  number={3},
  pages={273--297},
  year={1995},
  publisher={Springer}
}

@article{rumelhart1986learning,
  title={Learning Representations by Back-propagating Errors},
  author={Rumelhart, David E. and others},
  journal={Nature},
  volume={323},
  number={6088},
  pages={533--536},
  year={1986}
}

@misc{ruder2016overview,
  title={An Overview of Gradient Descent Optimization Algorithms},
  author={Ruder, Sebastian},
  year={2016},
  howpublished={\url{arXiv:1609.04747}}
}

@book{casella2002statistical,
  title={Statistical Inference},
  author={Casella, George and Berger, Roger L.},
  year={2024},
  edition={2},
  publisher={Chapman and Hall/CRC}
}

@book{boyd2004convex,
  title={Convex Optimization},
  author={Boyd, Stephen and Vandenberghe, Lieven},
  year={2004},
  publisher={Cambridge University Press}
}

@article{kingma2015adam,
  title={Adam: A Method for Stochastic Optimization},
  author={Kingma, Diederik P. and Ba, Jimmy},
  journal={ICLR},
  year={2015}
}

@inproceedings{sutskever2013importance,
  title={On the Importance of Initialization and Momentum in Deep Learning},
  author={Sutskever, Ilya and others},
  booktitle={ICML},
  pages={1139--1147},
  year={2013}
}

@book{nocedal2006numerical,
  title={Numerical Optimization},
 author={Wright, Stephen and others},
  year={1999},
  edition={2},
  publisher={Springer}
}

@book{norris1998markov,
  title={Markov Chains},
  author={Norris, J. R.},
  year={1998},
  publisher={Cambridge University Press}
}

@inproceedings{sulmont2019can,
  title={Can you teach me to machine learn?},
  author={Sulmont, Elisabeth and others},
  booktitle={Symposium on Computer Science Education},
  pages={948--954},
  year={2019}
}

@article{Sculley2015,
  author    = {Sculley, D. and others},
  title     = {Hidden Technical Debt in Machine Learning Systems},
  journal   = {NeurIPS},
  year      = {2015},
  volume    = {28},
  pages     = {2503--2511}
}

@online{gtAITechMinor,
  title        = {Minor in Applications of Artificial Intelligence and Machine Learning},
  organization = {Georgia Institute of Technology},
  year         = {2025},
  url          = {https://catalog.gatech.edu/programs/minor-artificial-intelligence-machine-learning/},
   note         = {Accessed: 2026-04-24}
}

@misc{duke,
  title        = {Minor in Machine Learning \& Artificial Intelligence},
  organization = {Duke University, Pratt School of Engineering, Department of Electrical \& Computer Engineering},
  year         = {2025},
  url          = {https://ece.duke.edu/academics/undergrad/minors/},
  note         = {Accessed: 2026-04-24}
}

@online{northwesternAIMinor,
  title        = {Artificial Intelligence Minor},
  organization = {Northwestern University, Department of Computer Science},
  year         = {2025},
  url          = {https://www.mccormick.northwestern.edu/computer-science/academics/undergraduate/ai-minor/} 
}

@book{Goodfellow-et-al-2016,
  title     = {Deep Learning},
  author    = {Ian Goodfellow and Yoshua Bengio and Aaron Courville},
  publisher = {MIT Press Cambridge},
  year      = {2016}
}

@book{Trefethen1997,
  title     = {Numerical Linear Algebra},
  author    = {Lloyd N. Trefethen and David Bau III},
  publisher = {SIAM},
  year      = {1997}
}

@article{Bastien2012,
  title={Theano: new features and speed improvements},
  author={Bastien, Fr{\'e}d{\'e}ric and others},
  journal={\url{arXiv:1211.5590}},
  year={2012}
}

@article{Bentley1975,
  author  = {Bentley, Jon Louis},
  title   = {Multidimensional Binary Search Trees Used for Associative Searching},
  journal = {Communications of the ACM},
  volume  = {18},
  number  = {9},
  year    = {1975},
  pages   = {509--517}
}

@article{Fredkin1960,
  author  = {Fredkin, Edward},
  title   = {Trie Memory},
  journal = {Communications of the ACM},
  year    = {1960},
  volume  = {3},
  number  = {9},
  pages   = {490--499}
}

@article{zha2025data,
  title={Data-centric artificial intelligence: A survey},
  author={Zha, Daochen and others},
  journal={ACM Computing Surveys},
  volume={57},
  number={5},
  pages={1--42},
  year={2025}
}

@article{Bloom1970,
  author  = {Bloom, Burton H.},
  title   = {Space/Time Trade-offs in Hash Coding with Allowable Errors},
  journal = {Communications of the ACM},
  year    = {1970},
  volume  = {13},
  number  = {7},
  pages   = {422--426}
}

@article{Pugh1990,
  author  = {Pugh, William},
  title   = {Skip Lists: A Probabilistic Alternative to Balanced Trees},
  journal = {Communications of the ACM},
  year    = {1990},
  volume  = {33},
  number  = {6},
  pages   = {668--676}
}

@book{Cormen2009,
  title     = {Introduction to Algorithms},
  author    = {Cormen, Thomas H. and others},
  edition   = {3rd},
  publisher = {MIT Press},
  year      = {2009}
}

@article{Robbins1951,
  author  = {Robbins, Herbert and Monro, Sutton},
  title   = {A Stochastic Approximation Method},
  journal = {Annals of Mathematical Statistics},
  volume  = {22},
  number  = {3},
  year    = {1951},
  pages   = {400--407}
}

@inproceedings{Bottou2010,
  author    = {Bottou, Léon},
  title     = {Large-Scale Machine Learning with Stochastic Gradient Descent},
  booktitle={Conf. on Computational Statistics},
  year      = {2010},
  pages     = {177--186}
}

@book{Holland1975,
  title     = {Adaptation in Natural and Artificial Systems},
  author    = {Holland, John H.},
  publisher = {University of Michigan Press},
  year      = {1975}
}

@article{Koren2009,
  author  = {Koren, Yehuda and others},
  title   = {Matrix Factorization Techniques for Recommender Systems},
  journal = {IEEE Computer},
  year    = {2009},
  volume  = {42},
  number  = {8},
  pages   = {30--37}
}

@inproceedings{Hinton2006,
  author    = {Hinton, Geoffrey E. and others},
  title     = {Reducing the Dimensionality of Data with Neural Networks},
  booktitle = {Science},
  volume    = {313},
  number    = {5786},
  pages     = {504--507},
  year      = {2006}
}

@article{Freund1997,
  author  = {Freund, Yoav and others},
  title   = {A Decision-Theoretic Generalization of On-Line Learning and an Application to Boosting},
  journal = {J. of Computer and System Sciences},
  volume  = {55},
  number  = {1},
  year    = {1997},
  pages   = {119--139}
}

@inproceedings{Chen2016,
  author    = {Chen, Tianqi and Guestrin, Carlos},
  title     = {{XGBoost}: A Scalable Tree Boosting System},
  booktitle = {Knowledge Discovery and Data Mining},
  year      = {2016},
  pages     = {785--794}
}

@inproceedings{Datar2002,
  author    = {Datar, Mayur and others},
  title     = {Maintaining Stream Statistics over Sliding Windows},
  booktitle = {SODA},
  year      = {2002},
  pages     = {635--644}
}

@article{Rosenfeld1966,
  author  = {Rosenfeld, Azriel and Pfaltz, John L.},
  title   = {Sequential Operations in Digital Picture Processing},
  journal = {J. of the ACM},
  year    = {1966},
  volume  = {13},
  number  = {4},
  pages   = {471--494}
}

@article{Codd1970,
  author  = {Codd, Edgar F.},
  title   = {A Relational Model of Data for Large Shared Data Banks},
  journal = {Communications of the ACM},
  volume  = {13},
  number  = {6},
  year    = {1970},
  pages   = {377--387}
}

@article{Chang2006,
   title={Bigtable: A distributed storage system for structured data},
  author={Chang, Fay and others},
  journal={TOCS},
  volume={26},
  number={2},
  pages={1--26},
  year={2008}
}

@book{Inmon1996,
  author    = {Inmon, William H.},
  title     = {Building the Data Warehouse},
  publisher = {John Wiley},
  year      = {1996}
}

@inproceedings{kahani2018automodel,
  title={AutoModel: a domain-Specific language for automatic modeling of real-time embedded systems},
  author={Kahani, Nafiseh},
  booktitle={ICSE: Companion Proceeedings},
  pages={515--517},
  year={2018}
}

@inproceedings{kahani2020bounded,
  title={Bounded verification of state machine models},
  author={Kahani, Nafiseh and Cordy, James R},
  booktitle={Proceedings of the 12th System Analysis and Modelling Conference},
  pages={23--32},
  year={2020}
}

@inproceedings{kahani2020synthesis,
  title={Synthesis of state machine models},
  author={Kahani, Nafiseh and others},
  booktitle={Model Driven Engineering Languages and Systems},
  pages={274--284},
  year={2020}
}

@article{Sawadogo2019,
   title={On Data Lake Architectures and Metadata Management},
  author={Sawadogo, Pegdwend{\'e} and others},
  journal={Intelligent Information Syst.},
  volume={56},
  number={1},
  pages={97--120},
  year={2021}
}

@inproceedings{Ghemawat2003,
  author    = {Ghemawat, Sanjay and orhers},
  title     = {The Google File System},  year      = {2003}
}

@inproceedings{Shvachko2010,
  title={The Hadoop Distributed File System},
  author={Shvachko, Konstantin and others},
  booktitle={MSST},
  pages={1--10},
  year={2010}}

@misc{Sanfilippo2010,
  author       = {Salvatore Sanfilippo},
  title        = {Redis: In-Memory Data Structure Store},
  url = {{https://redis.io/}}
 ,
  year         = {2025}
}

@inproceedings{Gonzalez2010,
    author={Low, Yucheng and others},
  title     = {GraphLab: A New Framework for Parallel Machine Learning},
  booktitle = {UAI},
  year      = {2010}
}

@article{Nickolls2008,
  title={Scalable parallel programming with {CUDA}: Is cuda the parallel programming model that application developers have been waiting for?},
  author={Nickolls, John  and others},
  journal={Queue},
  volume={6},
  number={2},
  pages={40--53},
  year={2008}}

@inproceedings{Jouppi2017,
   title={In-datacenter performance analysis of a tensor processing unit},
  author={Jouppi, Norman P and others},
  booktitle={Computer Architecture},
  pages={1--12},
  year={2017}
}

@inproceedings{Burns2016,
   author    = {Brendan Burns and others},
  title     = {Borg, Omega, and Kubernetes},
  booktitle = {Communications of the ACM},
  volume    = {59},
  number    = {5},
  year      = {2016},
  pages     = {50--57}
}

@inproceedings{Verma2015,
  title     = {Large-Scale Cluster Management at Google with Borg},
 author={Verma, Abhishek and others},
  booktitle={ 10th european  Conf. on computer systems},
  pages={1--17},
  year={2015}}

@inproceedings{Goldberg1974,
  author    = {Goldberg, Robert P.},
  title     = {Survey of Virtual Machine Research},
  booktitle = {IEEE Computer},
  year      = {1974}
}

@misc{Merkel2014,
  author       = {Merkel, Dirk},
  title        = {Docker: Lightweight Linux Containers for Consistent Development and Deployment},
  journal      = {Linux Journal},
  year         = {2014}
}

@inproceedings{Dean2008,
  author    = {Dean, Jeffrey and others},
  title     = {MapReduce: Simplified Data Processing on Large Clusters},
  booktitle = {Communications},
  volume    = {51},
  number    = {1},
  year      = {2008},
  pages     = {107--113}
}

@article{Zaharia2016,
  title     = {Apache Spark: A Unified Engine for Big Data Processing},
 author={Zaharia, Matei and  others},
  journal={Communications},
  volume={59},
  number={11},
  pages={56--65},
  year={2016}
}

@misc{XLA2017,
  author       = {Google},
  title        = {XLA: Optimizing Compiler for Machine Learning},
  year         = {2017},
  url = {https://www.tensorflow.org/xla}
}

@inproceedings{Chen2018,
  title={{TVM}: An automated {End-to-End} optimizing compiler for deep learning},
  author={Chen, Tianqi and  others},
  booktitle={OSDI 18},
  pages={578--594},
  year={2018}
}

@misc{Pfreundt2010,
  author    = {Hewlett-Packard Development Company, L.P.},
  title     = {Using InfiniBand for High Performance Computing},
  url       = {https://www.filibeto.org/unix/hp-ux/lib/os/infiniband/c00593119.pdf},
  year      = {2007}
}

@misc{NCCL2017,
  author       = {NVIDIA},
  title        = {NCCL: NVIDIA Collective Communications Library},
  year         = {2017},
  url = {{https://developer.nvidia.com/nccl}}
}

@inproceedings{MPI1994,
  author    = {Message Passing Interface Forum},
  title     = {MPI: A Message-Passing Interface Standard},
  booktitle = {Int. J. of Supercomputer Applications},
  volume    = {8},
  number    = {3-4},
  year      = {1994}
}

@inproceedings{Serban2020,
  title={Adoption and Effects of Software Engineering Best Practices in Machine Learning},
  author={Serban, Alex and others},
  booktitle={ESEM},
  pages={1--12},
  year={2020}
}

@inproceedings{Arpteg2018,
  title={Software engineering challenges of deep learning},
  author={Arpteg, Anders and others},
  booktitle={Soft. Eng. and Advanced Applications},
  pages={50--59},
  year={2018}
}

@article{Bosch2021EngineeringAI,
  title={Engineering {AI} Systems: A Research Agenda},
  author={Bosch, Jan and others},
  journal={IEEE Software},
  year={2021}
}

@techreport{Leslie2019,
  author = {Leslie, David},
  title = {Understanding Artificial Intelligence Ethics and Safety},
  institution = {The Alan Turing Institute},
  year = {2020}
}

@article{Mitchell2021,
  title        = {Model Cards for Model Reporting},
  author       = {Mitchell, Margaret and others},
  journal      = {10.48550/arXiv.1810.03993},
  Year      = {2020}
}

@article{serban2024software,
  title={Software Engineering Practices for Machine Learning—Adoption, Effects, and Team Assessment},
  author={Serban, Alex and others},
  journal={J. of Systems and Software},
  volume={209},
  pages={111907},
  year={2024}
}

@article{idowu2024machine,
  title={Machine Learning Experiment Management Tools: A Mixed-methods Empirical Study},
  author={Idowu, Samuel and others},
  journal={Empirical Soft. Eng.},
  volume={29},
  number={4},
  pages={74},
  year={2024}
}

@inproceedings{abdelkader2024ml,
  title={ML-On-Rails: Safeguarding Machine Learning Models in Software Systems—A Case Study},
  author={Abdelkader, Hala and others},
  booktitle={Conf. on AI Eng.-Soft. Eng. for AI},
  pages={178--183},
  year={2024}
}

@article{Kreuzberger2023MLOps,
  title={Machine Learning Operations (MLOps): Overview, Definition, and Architecture},
  author={Kreuzberger, Dominik and Kühl, Niklas and Hirschl, Sebastian},
  journal={IEEE Access},
  year={2023}
}

@article{Zhang2020,
  title={Review and empirical analysis of machine learning-based software effort estimation},
  author={Rahman, Mizanur and others},
  journal={IEEE Access},
  volume={12},
  pages={85661--85680},
  year={2024}
}

@article{Kim2022,
  title={Data scientists in software teams: State of the art and challenges},
  author={Kim, Miryung and others},
  journal={Trans. on Soft. Eng.},
  volume={44},
  number={11},
  pages={1024--1038},
  year={2017}
}

@inproceedings{cabral2024investigating,
  title={Investigating the Impact of Solid Design Principles on Machine Learning Code Understanding},
  author={Cabral, Raphael and others},
  booktitle={Conf. on AI Eng.-Soft. Eng. for AI},
  pages={7--17},
  year={2024}
}

@article{DoshiVelez2017,
  author = {Doshi-Velez, Finale and Kim, Been},
  title = {Towards A Rigorous Science of Interpretable Machine Learning},
  journal = {\url{arXiv:1702.08608}},
  year = {2017}
}

@article{Guidotti2018,
   author={Guidotti, Riccardo and others},
  title = {A Survey of Methods for Explaining Black Box Models},
  journal = {ACM Computing Surveys},
  volume = {51},
  number = {5},
  year = {2018}
}

@article{Polyzotis2018,
  title={Data lifecycle challenges in production machine learning: a survey},
  author={Polyzotis, Neoklis and others},
  journal={Sigmod Record},
  volume={47},
  number={2},
  pages={17--28},
  year={2018}}

@article{Schelter2018,
 title={Automating Large-scale Data Quality Verification},
  author={Schelter, Sebastian and Lange, Dustin and others},
  journal={ VLDB Endowment},
  volume={11},
  number={12},
  pages={1781--1794},
  year={2018}}

@article{Allamanis2018,
   title={A survey of machine learning for big code and naturalness},
  author={Allamanis, Miltiadis and others},
  journal={Computing Surveys},
  volume={51},
  number={4},
  pages={1--37},
  year={2018}
}

@article{Hindle2012,
  title={On the naturalness of software},
  author={Hindle, Abram and others},
  journal={Communications of the ACM},
  volume={59},
  number={5},
  pages={122--131},
  year={2016}
}

@article{Chen2021Codex,
   title={Evaluating Large Language Models Trained on Code},
  author={Chen, Mark and others},
  journal={\url{arXiv:2107.03374}},
  year={2021}
}

@article{FinnieAnsley2022,
  author = {Yeverechyahu, Doron and  Mayya, Raveesh and  Oestreicher-Singer, Gal },
  title = {The Impact of Large Language Models on Open-source Innovation: Evidence from GitHub Copilot},
 journal={\url{arXiv:2409.08379}},
  year = {2025}
}

@article{Ashmore2021,
  title={Assuring the machine learning lifecycle: Desiderata, methods, and challenges},
  author={Ashmore, Rob others},
  journal={CSUR},
  volume={54},
  number={5},
  pages={1--39},
  year={2021}
}

@inproceedings{Svyatkovskiy2020,
  title={Intellicode Compose: Code Generation using Transformer},
  author={Svyatkovskiy, Alexey and others},
  booktitle={ESEC/FSE},
  pages={1433--1443},
  year={2020}
}

@article{Ghaffarian2017,
  author = {Ghaffarian, Seyed Mohammad and Shahriari, Hamid Reza},
  title = {Software Vulnerability Analysis and Discovery Using Machine-Learning and Data-Mining Techniques: A Survey},
  journal = {ACM Computing Surveys},
  year = {2017}
}

@inproceedings{Just2014,
  title={{Defects4J}: A database of existing faults to enable controlled testing studies for Java programs},
  author={Just, Ren{\'e} and Jalali, Darioush and Ernst, Michael D},
  booktitle={ 2014 Int. symposium on software testing and analysis},
  pages={437--440},
  year={2014}
}

@article{Pradel2018,
  title={Deepbugs: A Learning Approach to Name-based Bug Detection},
  author={Pradel, Michael and others},
  journal={Prog. Lang.},
  volume={2},
  number={OOPSLA},
  pages={1--25},
  year={2018}
}

@article{Pan2019,
  title={Test case selection and prioritization using machine learning: a systematic literature review},
  author={Pan, Rongqi and others},
  journal={Empirical Soft. Eng.},
  volume={27},
  number={2},
  pages={29},
  year={2022},
  publisher={Springer}
}

@inproceedings{Breck2017,
  author = {Breck, Eric and others},
  title     = {The {ML} Test Score: A Rubric for {ML} Production Readiness and Technical Debt Reduction},
  booktitle = {Big Data},
  year      = {2017}

}

@article{ChenowethLinos2024,
  author  = {Steve Chenoweth and others},
  title   = {Teaching Machine Learning as Part of Agile Software Engineering},
  journal = {Education},
  volume  = {67},
  number  = {3},
  pages   = {377--386},
  year    = {2024}
}

@inproceedings{PereiraLGA2024OpenSourceLLMsCSEET,
  author    = {Juanan Pereira and others},
  title     = {Leveraging Open Source {LLM}s for Software Engineering Education and Training},
  booktitle = {CSEE\&T},
  year      = {2024},
  pages     = {1--10}}

@misc{KhanAkbarKasurinen2025LLMSEd,
  author       = {Maryam Khan and others},
  title        = {Integrating {LLM}s in Software Engineering Education: Motivators, Demotivators, and a Roadmap Towards a Framework for Finnish Higher Education Institutes},
  year         = {2025},
  howpublished = {arXiv:2503.22238},
  doi          = {10.48550/arXiv.2503.22238},
}

@misc{ChenEtAl2025PromptwareEngineering,
  author       = {Zhenpeng Chen and others},
  title        = {Promptware Engineering: Software Engineering for {LLM} Prompt Development},
  year         = {2025},
  howpublished = {arXiv:2503.02400},
  doi          = {10.48550/arXiv.2503.02400},
}

@misc{VillamizarEtAl2025PromptsArtifacts,
  author       = {Hugo Villamizar and others},
  title        = {Prompts as Software Engineering Artifacts: A Research Agenda and Preliminary Findings},
  year         = {2025},
  howpublished = {arXiv:2509.17548},
  doi          = {10.48550/arXiv.2509.17548},
}

@misc{equalexperts_mlops_playbook_v31_2022,
  author       = {Brabban, Paul and Carney, Jon and others},
  title        = {{MLOps} Playbook (v3.1)},
  howpublished = {Equal Experts Playbooks},
  year         = {2022},
  month        = may,
  url          = {https://www.equalexperts.com/wp-content/uploads/2022/05/MLOPS_Playbook_v3.1.pdf}
}

@misc{datatonic_mlops_guide_2021,
  author       = {{Datatonic}},
  title        = {The 2021 Guide to {MLOps}},
  howpublished = {Whitepaper},
  year         = {2021},
  url          = {https://cms.datatonic.com/app/uploads/2020/11/Datatonic-The-2021-Guide-to-MLOps-Whitepaper.pdf}
}

@misc{aws_mlops_checklist_2023,
  author       = {{Amazon Web Services (AWS)}},
  title        = {Evaluating your {ML} project with the {MLOps} checklist},
  howpublished = {AWS Prescriptive Guidance},
  year         = {2023}
}

@misc{googlecloud_mlops_cd_ct_2024,
  author       = {Kazmierczak, Jarek and others},
  title        = {{MLOps}: Continuous delivery and automation pipelines in machine learning},
  howpublished = {Google Cloud},
  year         = {2024}
}

@article{warnett2024understandability,
  title={On the understandability of MLOps system architectures},
  author={Warnett, Stephen John and Zdun, Uwe},
  journal={Soft. Eng.},
  volume={50},
  number={5},
  pages={1015--1039},
  year={2024},
  publisher={IEEE}
}

@inproceedings{obrien202223,
  title={23 shades of self-admitted technical debt: an empirical study on machine learning software},
  author={OBrien, David and  others},
  booktitle={European Soft.  Eng.  Conf.  on the Foundations of Soft. Eng.},
  pages={734--746},
  year={2022}
}

\end{document}